\documentstyle[11pt,newpasp,epsfig]{article}
\begin{document}
\title{Shaping the central structure in CDM halos}
\author{Amr El-Zant and Isaac Shlosman}
\affil{Department of Physics and Astronomy, University of
Kentucky, Lexington KY 40506, USA}
\author{Yehuda Hoffman}
\affil{Racah Institute of Physics, Hebrew University, Jerusalem, Israel}
\begin{abstract}
Coupling the dark matter (DM) and baryonic components in galaxies via dynamical
friction (by allowing the latter to be clumpy) leads to energy
input into the DM. The resulting expanding tendency in the DM component
overcomes the competing effect of adiabatic contraction, resulting from the
shrinking
gas component, and leads to a nearly constant density core. The  baryonic
mass inflow also leads to the coupling of the structure at the kpc scale
and below
and that of the structure of the DM at the 10 kpc scale and above, with rate of
mass inflow being dependent on the initial halo central concentration. The
latter
also determines the final halo core radius.

\end{abstract}

\section{Introduction}

The particles composing the halos to be studied here are theoretical
constructs; first imported from particle physics when it became clear
that (assuming general relativity is valid) most of the matter in the
Universe is invisible and that most of this material should be in
non-baryonic form (assuming standard big bang nucleosynthesis is
valid).
What distinguished the Cold Dark Matter (CDM) particles from the host of
other available options was their relatively small velocity dispersion
at the time of matter-radiation equality. 
This
meant that structure was not erased at the relevant scales.  It is
however precisely this property --- the particles being cold --- that
gave rise to the host of problems associated with CDM theory at the
kpc scale and below; dubbed the ``core catastrophe'' by Moore et al.
(1999).  These turn out to be mostly related to the large central
concentration of the resulting halos.

Possible solutions were again sought in particle physics; warm dark
matter and self-interacting dark matter being among the most popular.
The former however was found to be inadequate in washing out the cusp
(Colin, Avila-Rees \& Valenzuela 2000), while in the latter, in
general, the core catastrophe is replaced by a (more serious)
gravothermal catastrophe (Kochanek \& White 2000).  The basic idea
behind the interacting dark matter proposal is to transport energy to
the inner region, causing it to expand.  This does indeed happen,
because these systems initially have temperature inversion.  Once this
is washed away however, a runaway thermal instability develops ---
with the core contracting as it loses energy, while still getting
hotter (see Lynden Bell 1998 for a recent review of this phenomenon).

%YH: I suggest to add in the following paragraph a few sentences (to
%be taken from the papaer) on the (astro)physical motivation for
%assuming clumpiness of the gas and justification for the mass of the
%clumps. This is crucial for the paper. %YH

Recently, we have proposed a two component model with one component
(the CDM) always receiving energy and the other (baryonic) component
always losing it.  Moreover, we have shown that the CDM component's
tendency to expand overcomes the effect of adiabatic contraction.  The
coupling between the two components is provided by dynamical friction
--- with the mechanism being efficient over timescales of the order of
a Gyr, provided the Baryonic material is distributed in clumps of $>
0.01 \%$ the total mass of the system (El-Zant, Shlosman \& Hoffman
2001; ESH).  The resulting halos have rotation curves that are well fit by
those inferred from observations.  The final baryonic distribution
depends on the initial conditions (which determines the amount of
binding energy stored in that component) and the relative magnitude of
the dynamical friction to collision times --- with the latter
determining a natural termination of the process via collisions,
fragmentation, star formation etc.

As opposed to the dark matter, a centrally concentrated baryonic
density distribution is compatible with observations of galaxies ---
as in their bulge-black hole systems.  Baryonic material also
participates in electromagnetic and nuclear interactions, and
therefore can do things that CDM
%YH:
%can't
cannot %YH
(e.g., form stars and participate in wind driven outflows).  In this
context, processes taking place at the kpc scale and below are coupled
to the structure of dark matter at the $100$ kpc scale and, through
this, to the large scale structure of the Universe.  We will show in
this paper that there is a correlation between the rate of baryonic
mass inflow resulting from the coupling with the dark matter component
and the initial structure and subsequent evolution of the latter.
First we briefly discuss the method, initial conditions and values of
the parameters that are used.

\section{Method, parameters and initial conditions}

The method used here is
%YH:
%the same as
described %YH
in ESH: Baryonic clumps suffer
dynamical friction, described by the Chandrasekhar formula, and the
energy lost is redistributed among the dark matter particles using a
Monte Carlo technique.
%YH:
%A system of $100000$ particles is divided into a $1000$ equal number bins.
A system of $10^5$ particles is divided into a $10^3$ equal number
bins.  %YH
The energy lost by the gas particles of
mass $M$ and velocity ${\bf V}_M$ in a timestep $\Delta t$ in a given
bin
\begin{equation}
E_{lostbin} = \Sigma_{bin} M \left(\frac{ d \bf{V}_M} {dt} . {\bf V}_{M}
\right) \Delta t,
\end{equation}
is gained as random kinetic energy by
%YH:
%each component of
the %YH
halo particles in the same bin:
\begin{equation}
V_i \rightarrow V_i + F(y),
\end {equation}
where $F$ is a normal distribution of zero mean and variance
\begin{equation}
y = \sqrt{(2/3) E_{lostbin}/m_{bin}},
\end{equation}
and $m_{bin}$ represents the CDM mass in a given bin.
In the simulations presented here {\em both} the dark matter and baryonic
components
initially obey a Navarro, Frenk \& White (1997) (NFW) density distribution:
\begin{equation}
\rho_i = \frac{\rho_s}{r (r_s + r)^{2}}.
\label{NFW}
\end{equation}
The velocities are determined by solving the Jeans equation and are sampled
from a  Maxwellian distribution.
The baryonic mass fraction was fixed at $10 \%$ of the total
mass. The mass of a single baryonic lump was taken to be $0.1 \%$ the total
mass of
the system. The length and time units used take advantage of the
``universal'' (independent of mass)
scaling of the NFW halos and are given by
\begin{equation}
 [r]=1.63 M{_{DM}^{1/3}}  h^{-2/3} 10^{-4}\ {\rm kpc}
\end{equation}
and
\begin{equation}
[t]= 0.97 h^{-1} \sqrt{M_{DM}/M_{tot}}\ {\rm Myr}.
\end{equation}

\section{Results}

\begin{figure}
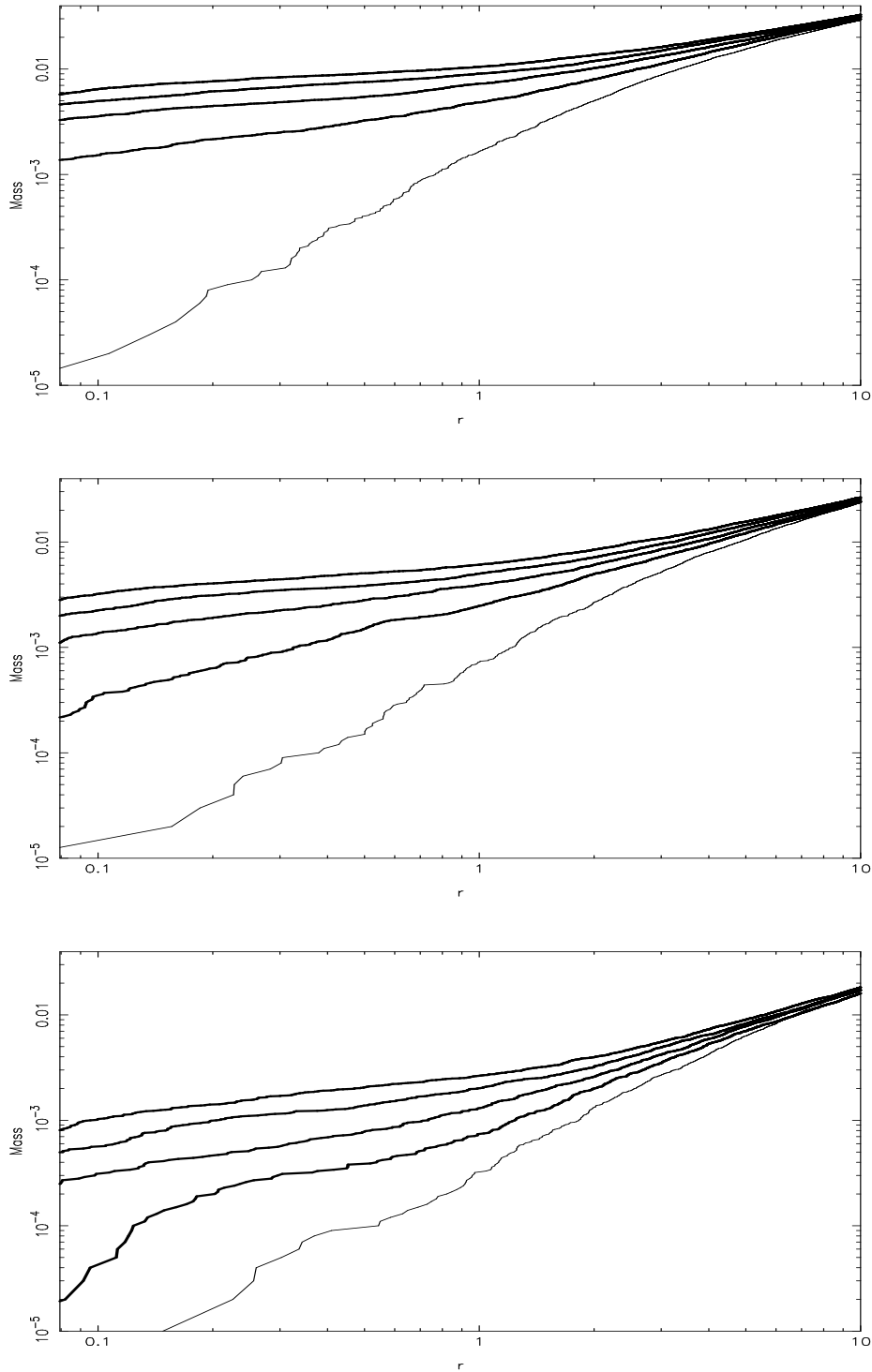

\plotfiddle{masses34.ps}{6.25cm}{-90}{50}{34}{-200pt}{+200pt}
\plotfiddle{masses33.ps}{6.25cm}{-90}{50}{34}{-200pt}{+200pt}
\plotfiddle{masses35.ps}{6.25cm}{-90}{50}{34}{-200pt}{+200pt}
%\begin{center}
\caption{The baryonic mass  distribution in systems with
(from top to bottom) $r_s = 3, 5, 10$ (Eq.~\ref{NFW}) at $T=0, 200, 400, 800$
(inflow results in monotonous mass increase within a given radius).
The total initial radius of the systems is  $100$
units. The total mass is taken as one unit, of which 0.1 is baryonic
}
\label{masses}
%\end{center}
\end{figure}

Fig.~\ref{masses} shows the time evolution of the baryonic mass
distribution for three values of
the NFW scaling parameter $r_s$ that roughly span the allowed range found
in simulations.
The evolution is much more rapid when $r_s$
is smaller: more {\em concentrated initial distributions evolve faster}.
This is a result of the higher central density --- leading  to stronger DF
coupling and
shorter dynamical time. It is also evident that by the final time shown
(corresponding
to about a Gyr) most of the  mass, originally within the inner $10$ kpc
or so, becomes concentrated near the center.
In reality this overestimates
the amount of mass inflow to the central region --- since we neglect here
effects due
to collisions, star formation etc., which would regulate the central influx
at high
densities. In practice then one would expect, in addition to a large
central mass component,
the process to result in a corresponding bulge component, the prominence of
which
would depend on the efficiency of the above mechanisms.
Different initial conditions for the baryonic component
 can also lead to very substantial
reduction in central concentration (compare, e.g.,
with top right panel of Fig.~1 of ESH).
 Nevertheless it is clear  that the processes described here can result in the
buildup of central mass concentration, and the formation of a bulge-black hole
component,  whose structure and properties are dependent on the initial
halo density
distibution.

\begin{figure}
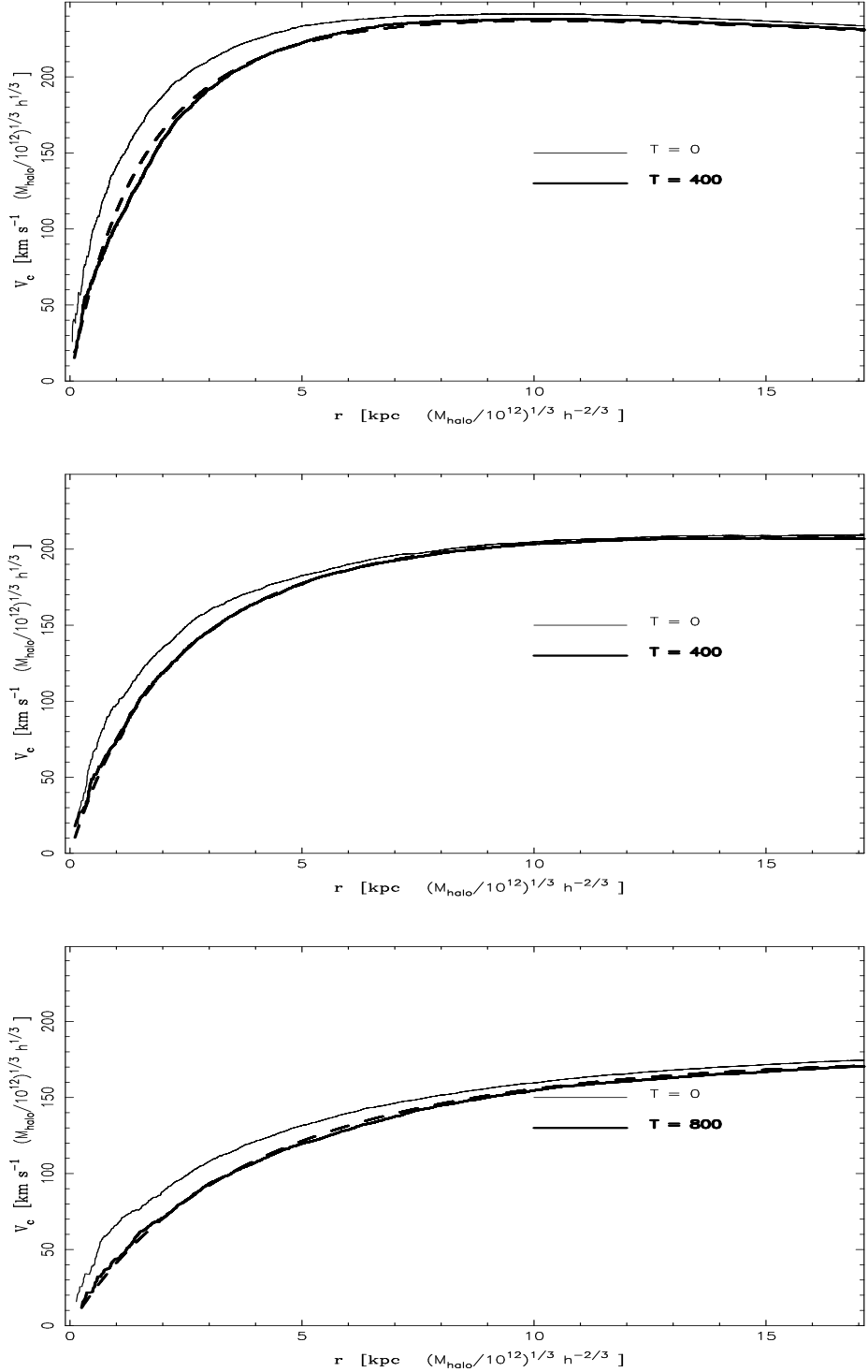

\plotfiddle{fits34.ps}{6.25cm}{-90}{50}{34}{-200pt}{+200pt}
\plotfiddle{fits33.ps}{6.25cm}{-90}{50}{34}{-200pt}{+200pt}
\plotfiddle{fits35.ps}{6.25cm}{-90}{50}{34}{-200pt}{+200pt}
%\begin{center}
\caption{Rotation curves of the CDM corresponding to systems of
Fig.~\ref{masses}.
Fits employ Eq.~(\ref{halfit}) with $A =1.44$, $r_c =1.43$ (top),
$A =2.11$, $r_c =2.10$ (middle) and  $A =3.52$, $r_c = 3.62$.
}
\label{fits}
%\end{center}
\end{figure}

Since the initial evolution proceeds ``inside out'' --- that is, with the
higher density regions
evolving at a faster rate --- the halo profile
is  first affected in the inner regions.  At this
stage, when a core develops in the very central region but the structure at
intermediate
radii is not affected, the density profile was found to be fit by
\begin{equation}
\rho = \frac{C}{(r_c+r)(A+r)^{2}}.
\label{halfit}
\end{equation}
The requirement that this equation fit the density at large radii fixes
$C$. As halos evolve,
this fit becomes a one-parameter fit, since it is found  that best fits
invariably have
$r_c \sim A$. We take here the point in time where this happens to define
the formation of a significant nearly constant density core. Since  outputs
are sampled
every $200$ time units, there is a certain coarse graining in our
determination of this time.
We find that this point is reached after $400$ time units (about half a
Gyr) for the cases of
$r_s = 3$ and $r_s = 5$ and double that time for case of $r_s=10$. The fits
are shown in
Fig.~\ref{fits}.
Note that the final core radii correlate with the initial $r_s$. As
explained in
ESH this can lead to interesting
correlations between
the initial profiles obtained from simulations and the final ones inferred from
observations.
As halos continue to evolve, their density structure at intermediate radii
is modified and
and is fit well by the
``Burkert profile'':
$\rho= \frac{\rho_0} { r_0^{2} + r^{2} }$. 
This happens within
$2 {\rm Gyr}$. The value of the Burkert core radii were also found to
correlate with the initial
NFW scale length. 
%For $r_s= 3, 5, 10$ we found $r_0=.....$. This is again
%not surprising,
%since the processes important for the evolution of our systems take place
%within the
%central region (where the density decreases as $1/r$ --- even though we
%have distributed our
%baryonic material throughout the CDM halo. Much of it does not participate).

\section{Concluding remarks}

The attempted solution of the apparent ``core catastrophe''
by  coupling the baryonic and dark matter components via dynamical friction,
 gives rise to interesting coupling between baryonic
 structural properties at the kpc scale and below and the dark matter
distribution
on the 10 kpc scale and above.
In this model, the  baryonic component is  clumpy,
%YH:
%contrary to what is assumed in most cosmological formation scenarios,
%but in line with ``viscous''formation models (e.g., Silk \& Norman 1981).
as is genrally expected during  galaxy formation
(e.g., Silk \& Norman 1981), but in contrast with many calculations
where a smooth baryonic component has been assumed for the sake of
simplicity. %YH
As a result, energy is pumped into the halo and it ``puffs up'', as the
expanding
tendency resulting from the {\em energy input overcomes the competing
effect of adiabatic
contraction}. At the same time,
the baryonic material loses energy and becomes more centrally concentrated.
This
material may participate in the formation of bulge-central mass component.
The details of this process are not studied here, since we do not model
effects of
collisions and star formation that will eventually dominate the evolution of
the clumpy gaseous component, and determine the relative efficiency of
formation of the
bulge and the central mass. Nevertheless, it is clear that
{\em the rate at which this  takes place depends on the dark matter halo's
initial
concentration --- with more concentrated halos being more efficient in fueling
central baryonic concentrations}.  The final halo core radius also depends
on the initial
halo scale length. Here we have used the NFW profile for the initial
conditions, whereas
the central density decreases as $1/r$, and the scale length determines a
turnoff radius
beyond which the density profile tends to $\rho \propto 1/r^{3}$. The
results presented
suggest that the mechanisms described above are likely to be more efficient
if CDM halos
are initially more concentrated than in the NFW picture (i.e., if the
central density
decreases as $\rho \propto 1/r^{\gamma}$, $\gamma > 1$ as inferred from
the higher resolution simulations).

\end{document}